\documentclass[prl,10pt,twocolumn,preprintnumbers,amsmath,amssymb,superscriptaddress,noshowpacs]{revtex4-1}
\usepackage{amsmath}
\usepackage{amssymb}
\usepackage{amsfonts}
\usepackage{eucal}
\usepackage[active]{srcltx}%
\usepackage{color}
\usepackage[all]{xy}
\usepackage{epsfig,graphicx,bm}
\usepackage[colorlinks=true, linkcolor=blue,urlcolor=magenta, filecolor=green,
citecolor=red,
pdfstartview=FitV,
pdftitle={},
pdfauthor={Kamal Hajian, Shahin Sheikh-Jabbari, Hossein Yavartanoo},
pdfsubject={},
pdfkeywords={},
pdfpagemode=None,
bookmarksopen=true
]{hyperref}
\usepackage{color}
\usepackage[normalem]{ulem}

\newcommand{\vacj}{|0;J_0\rangle}
\newcommand{\vac}{|0\rangle}
\newcommand{\nj}{|\{n_i\};{{J_0}}\rangle}

\newcommand{\HK}{{\cal H}_{Kerr}}
\newcommand{\HJ}{{\cal H}_{\bcJ}}
\newcommand{\vev}[1]{\langle{#1}\rangle}
\def\bW{\mathbb{W}}
\def\bJ{\mathbb{J}}
\def\bL{\mathbb{L}}

\newcommand{\bfL}{\boldsymbol{L}}

\newcommand{\bcJ}{{\boldsymbol{\mathcal{J}}}}

\def\mj{\mathrm{j}}

\newcommand{\ba}{\begin{array}}
\newcommand{\ea}{\end{array}}
\newcommand{\be}{\begin{equation}}
\newcommand{\ee}{\end{equation}}
\newcommand{\bea}{\begin{eqnarray}}
\newcommand{\eea}{\end{eqnarray}}
\newcommand{\bse}{\begin{subequations}}
\newcommand{\ese}{\end{subequations}}
\newcommand{\bi}{\begin{itemize}}
\newcommand{\ei}{\end{itemize}}

\definecolor{darkgreen}{rgb}{0,0.3,0}
\definecolor{darkblue}{rgb}{0,0,0.3}
\definecolor{darkred}{rgb}{0.7,0,0}

\newcommand{\old}[1]{}

\begin{document}

\preprint{
IPM/P-2017/037\cr}

{\vskip .2cm}
\title{Extreme Kerr black hole microstates with horizon fluff
}

\author{K.~Hajian}
\email{kamalhajian@ipm.ir}
\affiliation{School of Physics, Institute for Research in Fundamental Sciences (IPM),  P.O.Box 19395-5531, Tehran, Iran}

\author{M.M.~Sheikh-Jabbari}
\email{jabbari@theory.ipm.ac.ir}
\affiliation{School of Physics, Institute for Research in Fundamental Sciences (IPM),  P.O.Box 19395-5531, Tehran, Iran}

\author{H.~Yavartanoo}
\email{yavar@itp.ac.cn}
\affiliation{ State Key Laboratory of Theoretical Physics, Institute of Theoretical Physics, Chinese Academy of Sciences, Beijing 100190, China}

\begin{abstract}
We present a one-function family of solutions to 4D vacuum Einstein equations. While all diffeomorphic to the same extremal Kerr black hole, they are labeled by well-defined conserved charges and are hence distinct geometries. We show that this family of solutions forms a  phase space the symplectic structure of which is invariant under a $U(1)$ Kac-Moody algebra generated by currents $\mathbb{J}_n$ and Virasoro generators $\mathbb{L}_n$ with central charge six times angular momentum of the black hole. This symmetry algebra is well-defined everywhere in the spacetime, near the horizon or in the asymptotic flat region. Out of the appropriate combination of $\bJ_n$ charges, we construct another Virasoro algebra at the same central charge. Requiring that these two Virasoro algebras should describe the same system leads us to a proposal for identifying extreme Kerr black hole microstates, dubbed as extreme  Kerr fluff. Counting these microstates, we not only correctly reproduce the Bekenstein-Hawking entropy of extreme  Kerr black hole, but also its expected logarithmic corrections.
\end{abstract}


\maketitle


Existence of rotating black holes in the sky, besides the notable gravity wave detection by LIGO \cite{LIGO}, is backed by many different advanced x-ray astronomy observations \cite{X-ray}. The Kerr geometry \cite{Kerr-solution} provides a very good description of these black holes and is specified by two parameters, mass and angular momentum (spin). The spin of a Kerr black hole is theoretically bounded by its mass and the maximum possible angular momentum for a given mass happens for the so-called extremal black holes. There are now several observations confirming existence of (nearly) {extremal} black holes \cite{Extreme-Kerr-Observation}. 

Besides the observations, black holes pose many theoretical challenges. In particular, it is established that black holes should behave as thermodynamical systems with a given temperature and entropy \cite{BH-thermodynamics, Bekenstein, Hawking-radiation}.  The first step toward resolution of the black hole information paradox, see, e.g., \cite{information-loss}, may come from identifying ``black hole microstates,'' the statistical mechanical system underlying the thermodynamical behavior of black holes. {Extremal} black holes, on which we focus in this work, have vanishing temperature but, generically, a nonzero entropy. They are hence usually viewed as the simplest black holes to tackle the microstate problem.

The first successful example of black hole microstate counting was performed by Strominger and Vafa \cite{Strominger-Vafa}. Their proposal makes a heavy use of supersymmetry and the underlying ``quantum gravity'' structure, provided by string theory. This proposal has been extended to many other (nearly) supersymmetric black holes, all within string theory \cite{microstate-count-strings}. There are other proposals based on or inspired by the AdS/CFT, in particular, AdS$_3$/CFT$_2$, and using Cardy formula to account for the black hole entropy \cite{Strominger, Carlip, Kerr/CFT, Compere-Kerr-CFT-review}. This latter class is usually only apt for counting of microstates and not identifying them. Another idea for black hole microstate identification is the fuzzball proposal \cite{Fuzzball}, according which microstates of a black hole are  smooth, horizon-free geometries which are ``superposed'' to give rise to black holes as usual general relativity solutions.  Although at the level of idea, fuzzball proposal does not rely on supersymmetry, its explicit constructions so far, e.g. see \cite{Fuzzball-example} and references therein, crucially use supersymmetry.

In the statistical mechanical description of usual thermodynamical systems, however, we do not usually need to have a full quantum description of the system. One can argue based on the principle of decoupling of scales, that there is no reason why black holes should be different \cite{footnote 1}. On the other hand, there is strong uniqueness and ``no hair'' theorems \cite{no-hair} barring us from constructing black hole microstates within the strict reading of Einstein's equivalence principle (EEP). Nonetheless, geometries which are diffeomorphic to each other and hence equivalent under the strict EEP can be physically distinguishable if one can associate conserved charges to certain coordinate transformations relating these geometries; thereby leading to the notion of ``relaxed equivalence principle'' \cite{Residual-symmetry} and ``non-trivial diffeomorphisms.'' Prime examples of such geometries and their symmetries are the BMS algebra \cite{BMS} and the Brown-Henneaux analysis \cite{Brown-Henneaux}. This point of view has been used to nicely rederive  Weinberg's soft theorems \cite{soft-theorems}.

One may then hope {that} this set of geometries and the associated symmetry algebras can remedy the black hole microstate problem. In the last few years, there have been many papers, most notably \cite{HPS} where the term ``soft hair'' was coined, trying to formulate this idea. In \cite{BTZ-fluff}, see also \cite{Banados-fluff, AGSY}, building upon the analysis of \cite{NH-symmetries}, we presented the \emph{horizon fluff} proposal which realizes the idea of black hole microstate identification within the relaxed equivalence principle setting. The horizon fluff proposal has been worked out for three-dimensional black holes \cite{BTZ, Banados-geometries}. The two key points in the horizon fluff proposal are, (1) the notion of ``softness'' of modes is not the same from the near horizon or asymptotic observer viewpoints, and  {it} is the near horizon softness which is relevant to black hole microstates \cite{AGSY}, (2) the symmetry algebra labeling the nontrivial diffeomorphisms may have two complementary realizations, giving rise to a duality which is used to solve for microstates. In the 3D cases, this was argued to be a particle/(black) hole duality \cite{AGSY}.
In this work we show how the horizon fluff proposal works for {extremal} 4D Kerr.

\section{{Extremal Kerr Phase Space, Its Charge Algebra and Hilbert space}} The extremal Kerr black hole  (EKBH) metric in Boyer-Lindquist coordinate system is
\be
\begin{split}\label{Ext-Kerr-metric}
\mathrm{d}s^2=-&\frac{\Delta}{\Sigma}(\mathrm{d}t+\mathrm{m}\sin^2\theta\,\mathrm{d} \phi)^2+\frac{\Sigma}{\Delta}\mathrm{d}r^2+\Sigma\mathrm{d}\theta^2\\+& \frac{\sin^2\theta}{\Sigma}\left((r^2+\mathrm{m}^2)\mathrm{d}\phi+\mathrm{m}\mathrm{d}t\right)^2,
\end{split}\ee
where $\Delta=(r-{\mathrm{m}})^2$ and $\Sigma=r^2+\mathrm{m}^2\cos^2\theta$. 
The mass and angular momentum of EKBH are given by
\be 
{\mathrm M}=\frac{\mathrm{m}}{G_N},\quad\mathrm{j}=\frac{\mathrm{m}^2}{G_N},\quad 
\ee 
which saturates the extremality bound $\mathrm{j}\leq G_N\mathrm{M}^2$. For this metric the horizon is at $r=\mathrm{m}$, horizon area is $A_h=8\pi \mathrm{m}^2$ and hence the Bekenstein-Hawking entropy is
\be\label{B-H-entropy}
S_{\text{B.H.}}=\frac{A_h}{4G_N}=\frac{8\pi\mathrm{m}^2}{4G_N}=2\pi\mj.
\ee
We work in units where $\hbar=c=k_B=1$.
\paragraph{\textbf{Generating the phase space.}}
We construct the Extremal Kerr black hole phase space (EKPS)  by a simple shift,
\be\label{PS-generating}
\mathrm d\phi\rightarrow{J}(\phi)\mathrm{d}\phi,\qquad\frac{1}{2\pi}\int_0^{2\pi}J(\phi)d\phi=1, 
\ee
in the metric \eqref{Ext-Kerr-metric}. The condition on $J(\phi)$ in the phase space generating transformation \eqref{PS-generating} is to keep periodicity of $\phi$ coordinate $2\pi$. It is obvious that EKPS consists of geometries which are solutions to 4D vacuum Einstein equations and that all metrics in EKPS are black holes (have event and non-bifurcate, degenerate Killing horizons) the same as the metric \eqref{Ext-Kerr-metric}, with the same ADM mass and angular momentum. We shall discuss below how and in which sense these one-function family of geometries form a phase space.

\paragraph{\textbf{Symplectic symmetry generators, their charges and algebra.}}
The main new technical result in this letter is {the} presence of two vector fields
\be\label{hat-chi-eta}
\hat\chi=\epsilon(\phi)\partial_\phi,\qquad\eta=\frac{1}{2J(\phi)}\tilde{\epsilon}(\phi)\partial_\phi,
\ee
which are both nontrivial diffeomorphisms (have well-defined charge) over the EKPS. Here $\epsilon(\phi),\tilde{\epsilon}(\phi)$ are two arbitrary periodic functions of $\phi$. {A concise account of computations  establishing this result is presented in the Appendix}, here we only give an outline of the analysis. Recalling \eqref{PS-generating}, it is  seen that for $\tilde\epsilon=1$ vector field $\eta$ is the Killing vector over the whole EKPS whose associated conserved charge is  equal to {half of the} angular momentum $\mathrm{j}$. Note that the \emph{symmetry generators} $\hat\chi$ and $\eta$ are both along the $\partial_\phi$ direction and do not have any dependence on the rest of metric. This is in contrast to similar analysis for the near horizon extremal Kerr geometry \cite{Kerr/CFT, NHEG-algebra} and, among other things, allows us to define the symmetry generators and the corresponding charges at the horizon or in the asymptotic region. Moreover, $\hat\chi$ or $\eta$ can be used to move on this phase space, 
\be\label{hat-chi-metric-shift} 
\begin{split}
\delta_{\hat{\chi}}g_{\mu\nu}[J]&={\cal{L}}_{\hat\chi} g_{\mu\nu}=g_{\mu\nu}[J+\delta_{\hat{\chi}}J]-g_{\mu\nu}[J],\\
\delta_{\eta}g_{\mu\nu}[J]&={\cal{L}}_{\eta}g_{\mu\nu}=g_{\mu\nu}[J+\delta_{\eta}J]-g_{\mu\nu}[J],
\end{split}
\ee 
which yield
\be\label{delta-J}
\delta_{\hat{\chi}}J=(\epsilon J)',\qquad\delta_\eta J=\tilde{\epsilon}'/2,
\ee
to first order in $\epsilon$ and $\tilde\epsilon$.
The difference between the $\hat\chi$ and $\eta$ variations in  \eqref{delta-J}, as we will see below, leads to different charge algebras associated with $\hat\chi$ and $\eta$.

We should next show that \eqref{PS-generating} indeed generates a phase space. To this end, we use the covariant phase space method (CPSM) \cite{CPSM, Thesis-Compere-Seraj-Hajian}. We do not present the details of the analysis here, however, it is straightforward to verify that our one-function family of metrics form a phase space, the EKPS, in Einstein gravity with the Lee-Wald \cite{Lee-Wald} or Barnich-Brandt \cite{Barnich-Brandt} symplectic structures, and as our discussion around \eqref{hat-chi-metric-shift} shows, one can move over the phase space by the action of $\hat\chi$ and/or $\eta$. In other words, one should be able to associate well-defined conserved charges to $\hat\chi$ and $\eta$ which are generators of infinitesimal displacement on the phase space. These charges are functions of $J(\phi)$  and act on the phase space by the Poisson bracket defined by the  (Lee-Wald or Barnich-Brandt) symplectic structure \cite{Lee-Wald, Barnich-Brandt}.

Our other technical result which has important physical consequences is that these conserved charges are ``symplectic'' \cite{NHEG-algebra, symplectic, symplectic-AdS3}, meaning that they specify the symplecto-isometries of our phase space, i.e. they keep the symplectic two-form of the phase space intact. Physically this means that these charges may be defined by surface integrals over any compact two-surface at any value of $t,r$ coordinates \cite{Hajian:2015xlp}. Dealing with symplectic symmetries brings the important advantage that our charges may be defined at the horizon or in the asymptotic region or any radius in between. 

{To keep the discussions as nontechnical as possible here we only  present simple analysis yielding the charge algebra, details of the charge computation technicalities will be presented in the Appendix.} Let the charge variation associated with $\hat\chi[\epsilon],\eta[\tilde{\epsilon}]$ be respectively denoted by $\delta\hat{\mathbb{L}}[\epsilon]$ and $\delta \mathbb{J}[\tilde\epsilon]$. These charge variations, by construction, are linear in $\epsilon,\tilde\epsilon$ and a complete set of these charges may be obtained by taking them to be {$e^{in\phi},n\in\mathbb{Z}$}, for which we denote {$\hat\chi[e^{in\phi}]=\hat\chi_n,\eta[e^{in\phi}]=\eta_n$} and associated charge variations by $\delta\hat{\mathbb{L}}_n,\delta \mathbb{J}_n$.

The CPSM has two general results \cite{Thesis-Compere-Seraj-Hajian}:\\
(1) If the charge variations are integrable over the phase space (and we can hence talk about $\hat{\mathbb{L}}_n, \mathbb{J}_n$), then
\be\begin{split}
\delta_{\hat\chi_m}\hat{\mathbb{L}}_n=\{\hat{\mathbb{L}}_n,\hat{\mathbb{L}}_m\},&\qquad\delta_{\eta_m}\mathbb{J}_n=\{\mathbb{J}_n,\mathbb{J}_m\},\\
-\delta_{\hat\chi_m}{\mathbb{J}}_n=&\delta_{\eta_n}\hat{\mathbb{L}}_m=\{\hat{\mathbb{L}}_m,{\mathbb{J}}_n\},
\end{split}
\ee
where in the above $\hat\bL_n, \bJ_n$ are to be viewed as functions over the phase space and the bracket $\{,\}$ is the Poisson bracket on this phase space.\\ 
(2) Algebra (Poisson bracket) of charges, up to possible central terms, is the same as the algebra of corresponding generators. To state this explicitly, let us first recall  that
\be\label{CPSM-fund-theorm-2}\begin{split}
\{\hat\chi_m,\hat\chi_n\}_{\text{L.B.}}&=-i(m-n)\hat\chi_{m+n},\\
\{\eta_m,\eta_n\}_{\text{A.L.B.}}&=0,\\
\{\hat\chi_m,\eta_n\}_{\text{A.L.B.}}&=in\eta_{m+n},
\end{split}\ee 
where L.B. denotes Lie bracket and A.L.B. the ``adjusted Lie bracket,'' adjusted by the $J$-field dependence of the generators \cite{NHEG-algebra, symplectic-AdS3}
\be\label{adjusted-bracket}
\hspace*{-3mm}\{\zeta_m[J],\xi_n[J]\}_{\text{A.L.B.}}\equiv\{\zeta_m[J],\xi_n[J]\}_{\text{L.B.}}-\delta^J_{\zeta_m}\xi_n+\delta^J_{\xi_n}\zeta_m.\nonumber
\ee 
To stress the field dependence of $\zeta$ or $\xi$ vector fields, we have explicitly expressed them as $\zeta[J]$ or $\xi[J]$. Therefore, we have
\begin{align}\label{CPSM-fund-theorm-1}
\{\hat{\mathbb{L}}_m,\hat{\mathbb{L}}_n\}&=-i(m-n)\hat{\mathbb{L}}_{n+m}+{\text{up to central terms}},\nonumber\\ \{\mathbb{J}_m,\mathbb{J}_n\}&=0+{\text{up to central terms}},\\
\{\hat{\mathbb{L}}_m,{\mathbb{J}}_n\}&=in \mathbb{J}_{m+n}+{\text{up to central terms}}.\nonumber
\end{align}
That is, $\hat{\mathbb{L}}_n$ form the Witt or possibly Virasoro algebra while $\mathbb{{J}}_n$ are commuting or form a current (Heisenberg) algebra.
Our next task is to compute the central terms and also to specify the expression of the charges over the phase space, i.e. $\hat{\mathbb{L}}_n$ and $\mathbb{{J}}_n$ as a function of $J(\phi)$.

\paragraph{\textbf{Charges over the phase space.}} Standard CPSM analysis \cite{HSY-to-come} reveals that $\hat{\mathbb{L}}_n$ and $\mathbb{{J}}_n$ are 
integrable and
\be\label{Lm Ln commut}
\hspace*{-5mm}\{\hat{\mathbb{L}}_m,\hat{\mathbb{L}}_n\}=-\frac{\mathrm{j}}{2\pi }\int\mathrm{d}\phi\,e^{i(m+n)\phi}\left(2J(imJ+J')\right).
\ee 
Recalling \eqref{delta-J}, \eqref{CPSM-fund-theorm-2} and \eqref{CPSM-fund-theorm-1}, we learn that $\hat{\mathbb{L}}_n$ should satisfy a Witt algebra and hence
\be\label{Ln-J2}
\hat{\mathbb{L}}_n=\frac{\mathrm{j}}{2\pi}\int{\mathrm{d}}\phi\,e^{i n \phi}J^2.
\ee

The charges associated to $\eta_n$ can be evaluated in a similar fashion and we get
\be\label{JJ-J}
\mathbb{J}_n=\frac{\mathrm{j}}{2\pi}\int{\mathrm{d}}\phi\,e^{in\phi}J,
\ee 
with the $\mathbb{J}_n$ algebra
\begin{equation} \label{JJ comut rel}
\{\mathbb{J}_m,\mathbb{J}_n\}=i\frac{{n\mathrm{j}}}{4\pi}\int\mathrm{d}\phi\,e^{i(m+n)\phi}=\frac{in}{2}\mathrm{j}\delta_{m+n,0}.
\end{equation}
With the above, \eqref{Ln-J2} then yields
\be
\hat{\mathbb{L}}_n=\frac{1}{\mathrm j}\sum_{p}\mathbb{J}_{p}\mathbb{J}_{n-p},\quad\{\hat{\mathbb{L}}_m,\hat{\mathbb{L}}_n\}=-i(m-n)\hat{\mathbb{L}}_{m+n}
\ee
and as a consistency check, one may also show
$\{\hat{\mathbb{L}}_m,\mathbb{J}_n\}=in\mathbb{J}_{n+m}$,
in accord with \eqref{CPSM-fund-theorm-2}, \eqref{CPSM-fund-theorm-1}.

\paragraph{\textbf{Twisted Sugawara construction and the Virasoro algebra.}} Given the symplectic symmetry generators $\hat\chi, \eta$, any linear combination of them is also a symplectic symmetry generator. In particular, let us consider
{\be 
\chi[\epsilon(\phi)]\equiv\hat\chi[\epsilon]{+}\eta[\epsilon']=\Big(\epsilon{+}\frac{\epsilon'}{2J}\Big)\partial_{\phi}.
\ee} 
The charge associated with $\epsilon=e^{in\phi}$, $\mathbb{L}_n$, is then
\be\label{Virasoro-twisted-Sugawara}
\mathbb{L}_n=\hat{\mathbb{L}}_n{+}in \mathbb{J}_n
=\frac{1}{\mathrm j}\sum_p \mathbb{J}_p\mathbb{J}_{n-p}{+}in  \mathbb{J}_n
\ee 
and together with current $\mathbb{J}_n$ form a Kac-Moody algebra,
\be\label{Virasoro-PB}
\{\mathbb{L}_m, \mathbb{L}_n\}=-i(m-n)\mathbb{L}_{m+n}- i{m}^3 \frac{\mj}{2} \delta_{m+n,0}.
\ee
As expected and discussed, $\mathbb{J}_0$ is the charge corresponding to the Killing vector $\frac{1}{2J(\phi)} \partial_\phi$ and commutes with all the other generators of the algebra, $\mathbb{J}_0$ is the center element of the algebra.
\paragraph{\textbf{Quantizing the algebra.}} To quantize the charges, we assume that they are operators defined on a Hilbert space (which we construct below). To avoid cluttering we use the same notation for classical charges over the phase space and quantum (operator-valued) charges over the Hilbert space and denote both by $\bL_n, \bJ_n$. We can quantize the Poisson bracket of charges by replacing them with commutators,
\be\label{quantization}
\{,\}\to{-i[,]},
\ee
to obtain
\begin{align}\label{Kac-Moody}
[\mathbb{J}_m,\mathbb{J}_n]&=\frac{m}{2}{\mathrm j}\,\delta_{m+n,0},\nonumber\\
[\mathbb{L}_m,\mathbb{J}_n]&=-n\,\mathbb{J}_{m+n}-in^2\frac{\mathrm{j}}{2}\delta_{m+n,0},\\
[\mathbb{L}_m,\mathbb{L}_n]&=(m-n)\mathbb{L}_{m+n}+ {m}^3\frac{\mj}{2}\delta_{m+n,0}.\nonumber 
\end{align} 
{From the last equation, the central charge $c=6\mathrm{j}$ for the Virasoro algebra can be read. Nonetheless, there is a convention in our construction which (harmlessly) affects the magnitude of the central charge, as we describe here. The convention is the normalization of the generators $\eta_n \to \alpha \eta_n$. Accordingly, $\mathbb{J}_n\to \alpha \mathbb{J}_n$ with the new commutation relation 
\begin{equation} \label{alpha-J}
\{\mathbb{J}_m,\mathbb{J}_n\}=\frac{in\alpha^2}{2}\mathrm{j}\delta_{m+n,0}.
\end{equation}
Then, \eqref{Virasoro-twisted-Sugawara} would be rewritten as
\begin{equation}\label{alpha-L}
\mathbb{L}_n=\frac{1}{\alpha^2\mathrm j}\sum_p \mathbb{J}_p\mathbb{J}_{n-p}{+}in  \mathbb{J}_n
\end{equation}
which yields the redefined central charge $c\to \alpha^2 c$. Specifically, the choice of $\alpha^2 =2$ reproduces the Kerr/CFT central charge $c=12 \mathrm{j}$. However, as the reader will find in our later analysis, this convention will not affect our proposed microstate counting. } 
\paragraph{\textbf{Extremal Kerr Hilbert space.}} Given the algebra \eqref{Kac-Moody} one can construct Hilbert space of unitary representations of the Virasoro algebra ${\cal{H}}_{Kerr}$. To this end, we start with the vacuum state, $\vacj$ (see \cite{AGSY} for more detailed discussion),  
\be\label{vacj}
\mathbb{J}_0\vacj=\mj{J}_0\vacj,\qquad\mathbb{J}_n\vacj=0,\quad{n>0}, 
\ee
and take $\vacj$ to be normalized, $\langle J_0';0\vacj=\delta_{\mj',\mj}$. The other states in $\HK$ may then be constructed as
\be\label{n,j}
\hspace*{-5mm}\nj={\cal N}_{n_i}\prod_{n_i>0}\ \mathbb{J}_{-n_i}\vacj,\quad{\cal{N}}_{\{n_i\}}^{-2}=\prod n_i, 
\ee
where ${\cal N}_{\{n_i\}}$ is chosen such that  $\langle \mj';\{n'_i\}|\{n_i\};\mj\rangle=\delta_{\{n'_i\},\{n_i\}}\delta_{\mj',\mj}$. In \eqref{vacj} we have used the convention
compatible with \eqref{JJ-J}, i.e. the charge operators and the corresponding functions are related as 
\be
\vev{\mathbb{J}(\phi)}=\frac{c}{6}J(\phi),\qquad c=6\mj.
\ee
The geometries in the EKPS then come with $J_0=1$, \emph{cf.} \eqref{PS-generating}, while one may have cases with $J_0\neq 1$ corresponding to cases with deficits or excesses \cite{footnote 2}. For the vacuum states one may observe that 
$$\langle J_0';0|\mathbb{L}_n\vacj=\frac{c}{6}J_0^2\delta_{J_0',J_0}\delta_{n,0},$$
and hence geometries in EKPS have positive $L_0$ \cite{footnote 3}.

\paragraph{\textbf{Another construction for the Virasoro algebra.}} Positivity of norm condition for states in $\HK$, while requiring $\mathbb{J}_n^\dagger=\mathbb{J}_{-n},\ n\neq0$,  allows for both Hermitian and anti-Hermitian $\mathbb{J}_0$. In particular, $\HK$ includes states with imaginary $J_0=\pm i\nu/2$ with $\nu\in(0,1]$, which will be relevant to our microstate construction. {Moreover, noting $\bL, \bJ$ commutator in \eqref{Kac-Moody}, $\bJ$ is not a conformal primary operator,} see \cite{AGSY} for more details. To avoid dealing with anti-hermitian operators, one may instead introduce $W$ fields and $\bW$ operators \cite{AGSY},
\be
W(\phi)=e^{-2\int^\phi{J}},\qquad\bW=\; :\!\!e^{-\frac{2}{\mj}\int^\phi\bJ}\!\!:, 
\ee
where $::$ denotes normal ordering,  which have  ``twisted periodicity'' $W(\phi+2\pi)=e^{\pm 2\pi i\nu}W(\phi)$. In our setting $\mj$ is the angular momentum of the original Kerr black hole and it is expected to be quantized by Bohr-quantization. Therefore, central charge $c=6\mj$ is also integer-valued. Recalling spectral flow symmetry of the $U(1)$ Kac-Moody algebra and general expectations from quantum gravity \cite{Maldacena-Maoz, AGSY},  one expects $\nu$ to take $c$ discrete values, $\nu=r/c,\ r=1,2,\cdots, c$.  

{Using \eqref{Kac-Moody} one can see that $\bW$ is a conformal primary operator of weight one and we hence have $6\mj$ independent such primary fields $W^r(\phi)$}, each with different twisted periodicity. 
These fields provide a \emph{free field} representation for the Virasoro algebra at central charge $6\mj$.  For the details of the construction we refer the reader to analysis in section 4 of \cite{AGSY}.

This Virasoro algebra is more conveniently written in terms of $\bcJ_n$ 
operators which are the collection of Fourier modes of the $c=6\mj$ independent $W$-fields, $\bW^r_n$, into a single operator $\bcJ_n$ where $\bcJ_{pc+r}\propto \bW^r_p$ and,
\be
[\bcJ_m,\bcJ_n]=\frac{m}{2}\delta_{m+n,0}.
\ee
In terms of these operators, 
\be\label{Vir-j-bcJ}
\begin{split}
\bfL_n&=\frac{1}{6\mj}\sum_{m}:\!\!\bcJ_{6n\mj-m}\bcJ_m\!\!:,\\  [\bfL_m,\bfL_n]&=(m-n)\bfL_{m+n}+\frac{6\mj}{12}(m^3-m)\delta_{m+n,0}.
\end{split}\ee

To construct the Hilbert space for the above Virasoro algebra ${\cal H}_{\bcJ}$ we start with the vacuum state $|0\rangle$
\be
\bcJ_{n}|0\rangle=0,\qquad{n}\geq{0}.
\ee
The rest of states in $\HJ$ can be constructed as usual:
$$|\{n_i\}\rangle=\prod_{\{n_i>0\}} \bcJ_{-n_i}\vac.$$
One can readily see that for any $|\Psi\rangle\in \HJ$, $\bcJ_0|\Psi\rangle=0$. Since $\bcJ_0$ measures the energy from the near horizon viewpoint, $\HJ$ may be conveniently called Hilbert space of ``near horizon soft hairs.''

\section{Extremal Kerr Microstates}
We have given two different constructions for the same Virasoro algebra at central charge $6\mj$, in \eqref{Virasoro-twisted-Sugawara} and \eqref{Vir-j-bcJ} and the associated Hilbert spaces $\HK$ and $\HJ$. Recalling that $\bcJ$ was constructed from $\bW$ which in turn is constructed from $\bJ$, $\HJ$ and $\HK$ should be equivalent.  We can use this equivalence to identify microstates of extremal Kerr black hole. The analysis and arguments is as outlined and discussed for the 3D case of BTZ black holes, the horizon fluff proposal \cite{BTZ-fluff, Banados-fluff, AGSY} and will be discussed in more detail in \cite{HSY-to-come}.  To identify {extreme} Kerr fluff (microstates of extremal Kerr), we propose that these two Virasoros and the corresponding Hilbert spaces provide dual descriptions for the same physical system, i.e. we require $\bfL_n=\bL_n$, or more precisely,
\be\label{duality-proposal}
\frac{1}{6\mj}\sum_{m}:\!\!\bcJ_{6n\mj-m}\bcJ_{m}\!\!:\ =\frac{1}{\mj}\sum_m:\!\!\bJ_{n-m}\bJ_m\!\!:{+}in\bJ_n.
\ee
In $\HK$ the {extremal} Kerr black hole state is given by $|0;J_0=1\rangle$, or equivalently, $\vev{\bL_m}=\mj \delta_{m,0}$. This state  then corresponds to set of states $|{\cal B}(\{n_i\});\mj\rangle\in\HJ$, the extreme  Kerr fluff states, which satisfy 
\be\label{Fluff-equation}
\langle{\cal B}'(\{n_i\});\mj|\bfL_m|{\cal B}(\{n_i\});\mj\rangle=\mj\delta_{m,0}\delta_{{\cal B}',{\cal B}}.
\ee
The above, recalling \eqref{duality-proposal}, is evidently solved by
\be\label{Kerr-Fluff-states}
|{\cal{B}}(\{n_i\});\mj\rangle=|\{n_i\}\rangle,\qquad \sum_i n_i=6\mj^2.
\ee
{\paragraph{\textbf{Microstate counting, a consistency check.}}  For large $\mj$ number of states specified by \eqref{Kerr-Fluff-states} is given} by the standard Hardy-Ramanujan problem of number of ways $P_N$ a given integer $N$ (here $6\mj^2$) can be partitioned into non-negative integers (see \cite{Carlip-Hardy-Ramanujan} and references therein):  
\be\label{HR-counting}
P_N\simeq\frac{1}{4N\sqrt{3}}e^{2\pi\sqrt{\frac{N}{6}}},\qquad{N}\gg 1.
\ee
The logarithm of this number gives the black hole entropy 
\be\label{Entropy-count}
S(\mj)=2\pi\mj+\text{log-corrections},
\ee
reproducing the Bekenstein-Hawking entropy \eqref{B-H-entropy}.

{Had we introduced the $\alpha$ parameter through a normalization of $\bJ$, the central charge would change to $6\alpha^2 j$ and $\vev{L_0}$ to $j/\alpha^2$ \emph{cf.} \eqref{alpha-L}. Therefore, $N$ would remain unchanged. As expected, the entropy is independent of this normalization.}

\section{Discussion and outlook}

We showed how the horizon fluff idea can be worked through for the 4D {extremal} Kerr black hole. Our analysis expands upon  the Kerr/CFT analysis \cite{Kerr/CFT} in three important ways: (1) our symmetry algebra is defined over the whole {extremal} Kerr geometry and not only in the near horizon region; (2) we have introduced the {extremal} Kerr phase space, our symmetries are symplectic (and not just asymptotic) and, (3) besides the Virasoro, we have a current algebra and our symmetry generator diffeomorphisms are all along the azimuthal angle $\partial_\phi$. Our preliminary analysis shows that similar features can be extended to other extremal black holes in higher dimensions, in particular to the class discussed in \cite{NHEG-algebra}.

One obvious question which arises is whether similar analysis and horizon fluff proposal work for generic non-extremal Kerr geometry. The phase space corresponding to generic Kerr will presumably have two or four independent functions (rather than the single $J(\phi)$ in our analysis) \cite{BTZ-fluff} and consequently one expects to see a {larger} algebra than $U(1)$ Kac-Moody. This symmetry algebra is inevitably a subalgebra of the asymptotic BMS$_4$ symmetry \cite{BMS, BMS-Glenn}. Ideas and analysis discussed in  \cite{Compere-Long, DGGP} could be helpful in tackling this problem. 

The first check of our proposal was provided through reproducing the Bekenstein-Hawking area law. The non-trivial test, however, comes from the logarithmic corrections. The Hardy-Ramanujan counting \eqref{HR-counting} gives $S=2\pi\mj-2\ln\mj+\text{subleading}$. The Kerr/CFT analysis  the log-corrections for the 4D {extremal} case is not yet available \cite{log-Kerr/CFT}. Nonetheless, there are general analysis  by Sen \cite{log-sen} {which} divides the log-corrections into ``zero-mode'' and ``non-zero mode'' contributions. As discussed in Secs. 2 and 3 of \cite{log-sen}, a semiclassical analysis like ours is expected to only capture the zero-mode part. The nonzero mode part needs a more ``quantum gravity'' type treatment which in \cite{log-sen} was performed using quantum entropy function and Euclidean quantum gravity approach. The zero-mode part of \cite{log-sen} matches with our result of $-2\ln \mj$. 

{We stated and used a ``duality'' between $\bcJ_n$ and $\bJ_n$  and the corresponding Hilbert spaces and discussed that in large $\mj$ limit the $\bcJ_n$  provide a free field representation for classical extreme Kerr microstates. Addressing questions of great interest like black hole evaporation dynamics and information paradox requires turning on microstate-microstates or microstate-background interactions, which are $1/\mj$ effects we did not consider here. Nonetheless, there are interesting potentially observable effects associated with  energy and angular momentum distribution of the microstates. Our proposal, as seen from \eqref{Fluff-equation}, \eqref{Kerr-Fluff-states}, has a specific spectrum.  We intend to explore such possible observable effects in our upcoming studies.}

\begin{acknowledgments}
\textbf{Acknowledgments:} We especially thank Hamid Afshar and Daniel Grumiller for their role in the development of the \emph{horizon fluff} proposal and for comments and discussions. We also thank Geoffrey Compere, Leo Pando Zayas, Maurice van Putten and Andrew Strominger for discussions and comments on the draft. H.Y. is supported in part by National Natural Science Foundation of China, Project 11675244.  M.M.Sh-J is supported in part by junior research chair in black hole physics of  Iranian NSF, Grant No. 950124 and the ICTP Simons associates fellowship. 
\end{acknowledgments}

\appendix
\section{Appendix: Details of charge calculations}
Here we provide more details about the conserved charge calculations presented in this paper. For clarity, we focus on the general relativity which is the gravitational theory in our analysis. In CPSM \cite{Lee-Wald}, conserved charge variations $\delta H$ in  general relativity $\mathcal{L}=\frac{1}{16\pi G_N}R$  in $4$ dimensions are calculated by a 2-form $\boldsymbol{k}_\xi (g_{\mu\nu}, \delta g_{\alpha\beta})$ integrated over asymptotic two dimensional boundary (which may be taken to be the two sphere at infinity)
\begin{equation}
\delta H_\xi = \oint_{S^2} \boldsymbol{k}_\xi (g_{\alpha\beta}, \delta g_{\alpha\beta}). 
\end{equation}
The  $\boldsymbol{k}$ is determined by Lagrangian (find the details of how to find the $\boldsymbol{k}$ from a Lagrangian e.g. in \cite{Hajian:2016kxx,Ghodrati:2016vvf}) to be Hodge dual to 
\begin{align}
k_\xi^{\mu\nu}(\delta g_{\alpha\beta},g_{\alpha\beta})&=\dfrac{1}{16 \pi G}\Big(\Big[\xi^\nu\nabla^\mu h
-\xi^\nu\nabla_\tau h^{\mu\tau}+\xi_\tau\nabla^{\nu}h^{\mu\tau}\nonumber\\
&+\frac{1}{2}h\nabla^{\nu} \xi^{\mu}-h^{\tau\nu}\nabla_\tau\xi^{\mu}\Big]-[\mu\leftrightarrow\nu]\Big),\label{k EH integrand}
\end{align} 
where $h_{\mu\nu}\equiv \delta g_{\mu \nu}$, $h\equiv h^\mu_{\,\,\mu}$ and $\epsilon_{\mu\nu\sigma\rho}$ is the Levi-Civita symbol such that $\epsilon_{tr\theta\varphi}=+1$. 
As the notation suggests, $\boldsymbol{k}$ has three inputs: (1) the background solution $g_{\alpha\beta}$, (2) linearized field perturbations $\delta g_{\alpha\beta}$, (3) symmetry generator $\xi$. For the symmetries generated by diffeomorphisms, $\xi$ is a vector field $\xi^\mu$. The first two inputs are determined by the phase space under consideration. In our analysis in this paper, the metric $g_{\alpha\beta}$ is the extremal Kerr black hole \eqref{Ext-Kerr-metric} deformed by the transformations \eqref{PS-generating}, i.e. $g_{\alpha\beta}[J]$. The perturbations $\delta g_{\alpha\beta}$ are chosen from the tangent of the phase space, which are either $\mathcal{L}_\eta g_{\alpha\beta}[J]$ or $\mathcal{L}_{\hat \chi} g_{\alpha\beta}[J]$, or any linear combination of them. The vectors $\xi^\mu$ for which we calculated charges over the phase space could be also chosen to be either $\eta^\mu$ or $\hat{\chi}^\mu$ or any linear combination of them. In summary, charge variations in this paper are explicitly calculated by:
\begin{align}
&-\delta_{\hat\chi_m} \hat{\mathbb{L}}_n=\delta_{\hat\chi_m} H_{\hat{\chi_n}} = \oint_\infty \boldsymbol{k}_{\hat\chi_n} (g_{\alpha\beta}, \delta_{\hat\chi_m} g_{\alpha\beta}),\nonumber\\
&-\delta_{\hat\chi_m} {\mathbb{J}}_n=\delta_{\hat{\chi}_m} H_{\eta_n} = \oint_\infty \boldsymbol{k}_{\eta_n} (g_{\alpha\beta}, \delta_{\hat{\chi}_m} g_{\alpha\beta}),\\
&-\delta_{\eta_m} \hat{\mathbb{J}}_n=\delta_{\eta_m} H_{\eta_n} = \oint_\infty \boldsymbol{k}_{\eta_n} (g_{\alpha\beta}, \delta_{\eta_m} g_{\alpha\beta}),\nonumber
\end{align}      
in which we considered the conventional minus sign (in the same footing as the angular momentum) in defining the charges $\hat{\mathbb{L}_n}$ and $\mathbb{J}_n$ with respect to the $H_{\chi_n}$ and $H_{\eta_n}$. 

It can be checked that for all of the $\boldsymbol{k}$'s in the relations above we have $d\boldsymbol{k}=0$. So, $\eta_n$ and $\hat{\chi}_n$ (and hence, any linear combination of them) are symplectic symmetries over the proposed phase space. Therefore by the Stokes' theorem, the integration $\oint_{S^2}$ can be relaxed to be taken over any closed surface  which is a smooth deformation of the boundary at infinity into the bulk. The rest of the analysis would be just performing the calculations. The results turn out to be
\begin{align}
&\delta_{\hat\chi_m} \hat{\mathbb{L}}_n=\frac{\mathrm{j}}{2\pi}\int_0^{2\pi} d \phi \, e^{i(m+n)\phi} \left(2J(imJ+J')\right),\label{hat L n explicit}\\
& \delta_{\hat\chi_m} {\mathbb{J}}_n=\frac{\mathrm{j}}{2\pi}\int_0^{2\pi} d \phi \, e^{i(m+n)\phi} (imJ+J'),\label{J explicit}\\
&\delta_{\eta_m} {\mathbb{J}}_n=i\frac{\mathrm{nj}}{4\pi}\int_0^{2\pi} d \phi \, e^{i(m+n)\phi} .\label{J J explicit}
\end{align} 
Considering the relation \eqref{delta-J}, i.e. $\delta_{\hat{\chi}_m}J=(e^{im\phi} J)'=e^{im\phi}(imJ+J')$, from equations \eqref{hat L n explicit} and \eqref{J explicit} we can simply read    
\begin{align}
\hat{\mathbb{L}}_n=\frac{\mathrm{j}}{2\pi}\int_0^{2\pi} d \phi \, e^{in\phi} J^2, \qquad {\mathbb{J}}_n=\frac{\mathrm{j}}{2\pi}\int_0^{2\pi} d \phi \, e^{in\phi} J.
\end{align}
Besides, the Eq. \eqref{J J explicit} is nothing but the commutation relation \eqref{JJ comut rel}.


\begin{thebibliography}{99}%


\bibitem{LIGO} 
B.~P.~Abbott {\it et al.} [LIGO Scientific and Virgo Collaborations],
Phys.\ Rev.\ Lett.\  {\bf 116}, no. 6, 061102 (2016)
\href{http://arxiv.org/abs/1602.03837}{[arXiv:1602.03837 [gr-qc]]}; 
Phys.\ Rev.\ Lett.\  {\bf 116}, no. 24, 241103 (2016)
\href{http://arxiv.org/abs/1606.04855}{[arXiv:1606.04855 [gr-qc]]}.
\bibitem{X-ray} 
H.~Tananbaum, M.~C.~Weisskopf, W.~Tucker, B.~Wilkes and P.~Edmonds,
Rept.\ Prog.\ Phys.\  {\bf 77}, 066902 (2014)
\href{http://arxiv.org/abs/1405.7847}{[arXiv:1405.7847 [astro-ph.HE]]}, and references therein.


\bibitem{Kerr-solution} 
R.~P.~Kerr,
Phys.\ Rev.\ Lett.\  {\bf 11}, 237 (1963).


\bibitem{Extreme-Kerr-Observation} 
J.~E.~McClintock, R.~Shafee, R.~Narayan, R.~A.~Remillard, S.~W.~Davis and L.~X.~Li,
Astrophys.\ J.\  {\bf 652}, 518 (2006)
\href{http://arxiv.org/abs/astro-ph/0606076}{[astro-ph/0606076]} $\bullet$

L.~Gou {\it et al.},
Astrophys.\ J.\  {\bf 790}, no. 1, 29 (2014)
\href{http://arxiv.org/abs/1308.4760}{[arXiv:1308.4760 [astro-ph.HE]]}.

\bibitem{BH-thermodynamics}
J.~M.~Bardeen, B.~Carter and S.~W.~Hawking,
Commun.\ Math.\ Phys.\  {\bf 31}, 161 (1973).

\bibitem{Bekenstein}
J.~D.~Bekenstein,
Lett.\ Nuovo Cim.\  {\bf 4}, 737 (1972);


Phys.\ Rev.\ D {\bf 7}, 2333 (1973) $\bullet$

J. D. Bekenstein, Phys. Rev. D 7, 2333 (1973).
\bibitem{Hawking-radiation}
S.~W.~Hawking,
Commun.\ Math.\ Phys.\  {\bf 43}, 199 (1975)
Erratum: [Commun.\ Math.\ Phys.\  {\bf 46}, 206 (1976)].


\bibitem{information-loss}
W.~G.~Unruh and R.~M.~Wald,
Rept.\ Prog.\ Phys.\  {\bf 80}, no. 9, 092002 (2017)
\href{http://arxiv.org/abs/1703.02140}{[arXiv:1703.02140 [hep-th]]} $\bullet$
S.~L.~Braunstein, S.~Pirandola and K.~Życzkowski,
Phys.\ Rev.\ Lett.\  {\bf 110}, no. 10, 101301 (2013)
\href{http://arxiv.org/abs/0907.1190}{[arXiv:0907.1190 [quant-ph]]} $\bullet$
S.~D.~Mathur,
Class.\ Quant.\ Grav.\  {\bf 26}, 224001 (2009)
\href{http://arxiv.org/abs/0909.1038}{[arXiv:0909.1038 [hep-th]]}.


\bibitem{Strominger-Vafa} 
A.~Strominger and C.~Vafa,
Phys.\ Lett.\ B {\bf 379}, 99 (1996)
\href{http://arxiv.org/abs/hep-th/9601029}{[hep-th/9601029]}.


\bibitem{microstate-count-strings} 
K.~Sfetsos and K.~Skenderis,
Nucl.\ Phys.\ B {\bf 517}, 179 (1998)
\href{http://arxiv.org/abs/hep-th/9711138}{[hep-th/9711138]}  $\bullet$ J.~R.~David, G.~Mandal and S.~R.~Wadia,
Phys.\ Rept.\  {\bf 369}, 549 (2002)
\href{http://arxiv.org/abs/hep-th/0203048}{[hep-th/0203048]} $\bullet$
A.~Dabholkar,
Phys.\ Rev.\ Lett.\  {\bf 94}, 241301 (2005)
\href{http://arxiv.org/abs/hep-th/0409148}{[hep-th/0409148]} $\bullet$
A.~Sen,
Gen.\ Rel.\ Grav.\  {\bf 40}, 2249 (2008)
\href{http://arxiv.org/abs/0708.1270}{[arXiv:0708.1270 [hep-th]]}.

\bibitem{Strominger} 
A.~Strominger,
JHEP {\bf 9802}, 009 (1998)
\href{http://arxiv.org/abs/hep-th/9712251}{[hep-th/9712251]}.


\bibitem{Carlip} 
S.~Carlip,
Class.\ Quant.\ Grav.\  {\bf 16}, 3327 (1999)
\href{http://arxiv.org/abs/gr-qc/9906126}{[gr-qc/9906126]};
Phys.\ Rev.\ Lett.\  {\bf 82}, 2828 (1999)
\href{http://arxiv.org/abs/hep-th/9812013}{[hep-th/9812013]};
Phys.\ Rev.\ Lett.\  {\bf 88}, 241301 (2002)
\href{http://arxiv.org/abs/gr-qc/0203001}{[gr-qc/0203001]}.


\bibitem{Kerr/CFT} 
M.~Guica, T.~Hartman, W.~Song, A.~Strominger,
Phys.\ Rev.\ D {\bf 80}, 124008 (2009)
\href{http://arxiv.org/abs/0809.4266}{[arXiv:0809.4266 [hep-th]]}.

\bibitem{Compere-Kerr-CFT-review}
G.~Comp\`ere,
Living Rev.\ Rel.\  {\bf 15}, 11 (2012)
\href{http://arxiv.org/abs/1203.3561}{[arXiv:1203.3561 [hep-th]]}.


\bibitem{Fuzzball}
S.~D.~Mathur,
\href{http://arxiv.org/abs/0810.4525}{arXiv:0810.4525 [hep-th]};
Fortsch.\ Phys.\  {\bf 53}, 793 (2005)
\href{http://arxiv.org/abs/hep-th/0502050}{[hep-th/0502050]}.


\bibitem{Fuzzball-example} 
I.~Bena, S.~Giusto, E.~J.~Martinec, R.~Russo, M.~Shigemori, D.~Turton and N.~P.~Warner,
Phys.\ Rev.\ Lett.\  {\bf 117}, no. 20, 201601 (2016)
\href{http://arxiv.org/abs/1607.03908}{[arXiv:1607.03908 [hep-th]]}.

\bibitem{footnote 1}
{The black hole complementarity principle \cite{BH-complementarity} is a nice expression of this idea. It is, however, facing problems with firewall controversy \cite{firewall}. }

\bibitem{BH-complementarity}
L.~Susskind, L.~Thorlacius and J.~Uglum,
Phys.\ Rev.\ D {\bf 48}, 3743 (1993)
\href{http://arxiv.org/abs/hep-th/9306069}{[hep-th/9306069]}.
\bibitem{firewall}
A.~Almheiri, D.~Marolf, J.~Polchinski and J.~Sully,
JHEP {\bf 1302}, 062 (2013)
\href{http://arxiv.org/abs/1207.3123}{[arXiv:1207.3123 [hep-th]]}.

\bibitem{no-hair} 
R.~Ruffini and J.~A.~Wheeler,
Phys.\ Today {\bf 24}, no. 1, 30 (1971) $\bullet$
P.~T.~Chrusciel,
Contemp.\ Math.\  {\bf 170}, 23 (1994)
\href{http://arxiv.org/abs/gr-qc/9402032}{[gr-qc/9402032]} $\bullet$
J.~D.~Bekenstein,
``Black hole hair: 25 - years after,''
In *Moscow 1996, 2nd International A.D. Sakharov Conference on physics* 216-219
\href{http://arxiv.org/abs/gr-qc/9605059}{[gr-qc/9605059]}.


\bibitem{Residual-symmetry} 
M.~M.~Sheikh-Jabbari,
Int.\ J.\ Mod.\ Phys.\ D {\bf 25}, no. 12, 1644019 (2016)
\href{http://arxiv.org/abs/1603.07862}{[arXiv:1603.07862 [hep-th]]} and
seminar presented in workshop on \emph{Recent developments in symmetries and (super)gravity theories}, June 2016,  Bogazici Uni. Istanbul.


\bibitem{BMS} 
H.~Bondi, M.~G.~J.~van der Burg and A.~W.~K.~Metzner,
Proc.\ Roy.\ Soc.\ Lond.\ A {\bf 269}, 21 (1962) $\bullet$
R.~K.~Sachs,
Proc.\ Roy.\ Soc.\ Lond.\ A {\bf 270}, 103 (1962);  
Phys.\ Rev.\  {\bf 128}, 2851 (1962).


\bibitem{Brown-Henneaux}
J. D. Brown and M. Henneaux, 
Commun. Math. Phys. \textbf{104}, 207 (1986).

\bibitem{soft-theorems}
T.~He, V.~Lysov, P.~Mitra and A.~Strominger,
JHEP {\bf 1505}, 151 (2015)
\href{http://arxiv.org/abs/1401.7026}{[arXiv:1401.7026 [hep-th]]} $\bullet$ V.~Lysov, S.~Pasterski and A.~Strominger,
Phys.\ Rev.\ Lett.\  {\bf 113}, no. 11, 111601 (2014)
\href{http://arxiv.org/abs/1407.3814}{[arXiv:1407.3814 [hep-th]]} $\bullet$ For a comprehensive review, see A.~Strominger,
\href{http://arxiv.org/abs/1703.05448}{arXiv:1703.05448 [hep-th]}.


\bibitem{HPS} 
S.~W.~Hawking, M.~J.~Perry and A.~Strominger,
Phys.\ Rev.\ Lett.\  {\bf 116}, no. 23, 231301 (2016)
\href{http://arxiv.org/abs/1601.00921}{[arXiv:1601.00921 [hep-th]]};  JHEP {\bf 1705}, 161 (2017)
\href{http://arxiv.org/abs/1611.09175}{[arXiv:1611.09175 [hep-th]]}.


\bibitem{BTZ-fluff} 
H.~Afshar, D.~Grumiller and M.~M.~Sheikh-Jabbari,
Phys.\ Rev.\ D {\bf 96}, 96 (2017) no.8, 084032, \href{http://arxiv.org/abs/1607.00009}{[arXiv:1607.00009 [hep-th]]}.


\bibitem{Banados-fluff}
M.~M.~Sheikh-Jabbari and H.~Yavartanoo,
Phys.\ Rev.\ D {\bf 95}, no. 4, 044007 (2017)
\href{http://arxiv.org/abs/1608.01293}{[arXiv:1608.01293 [hep-th]]}.
\bibitem{AGSY} 
H.~Afshar, D.~Grumiller, M.~M.~Sheikh-Jabbari and H.~Yavartanoo,
JHEP {\bf 1708}, 087 (2017) \href{http://arxiv.org/abs/1705.06257}{[arXiv:1705.06257 [hep-th]]}. 


\bibitem{NH-symmetries} 
H.~Afshar, S.~Detournay, D.~Grumiller, W.~Merbis, A.~Perez, D.~Tempo and R.~Troncoso,
Phys.\ Rev.\ D {\bf 93}, no. 10, 101503 (2016)
\href{http://arxiv.org/abs/1603.04824}{[arXiv:1603.04824 [hep-th]]} $\bullet$
H.~Afshar, D.~Grumiller, W.~Merbis, A.~Perez, D.~Tempo and R.~Troncoso,
Phys.\ Rev.\ D {\bf 95}, no. 10, 106005 (2017)
\href{http://arxiv.org/abs/1611.09783}{[arXiv:1611.09783 [hep-th]]}.


\bibitem{BTZ}
M.~Ba\~nados, C.~Teitelboim and J.~Zanelli, 
Phys.\ Rev.\ Lett.\  {\bf 69}, 1849 (1992),
\href{http://arxiv.org/abs/hep-th/9204099}{[hep-th/9204099]}.

\bibitem{Banados-geometries}
M.~{Ba{\~n}ados}, AIP Conf.Proc. 484 (1999) 147
\href{http://arxiv.org/abs/hep-th/9901148}{{[hep-th/9901148]}}.


\bibitem{NHEG-algebra} 
G.~Comp\`ere, K.~Hajian, A.~Seraj and M.~M.~Sheikh-Jabbari,
Phys.\ Lett.\ B {\bf 749}, 443 (2015),  \href{http://arxiv.org/abs/1503.07861}{[arXiv:1503.07861 [hep-th]]};
JHEP {\bf 1510}, 093 (2015)
\href{http://arxiv.org/abs/1506.07181}{[arXiv:1506.07181 [hep-th]]}.

\bibitem{CPSM} 
C.~Crnkovic and E.~Witten,
In *Hawking, S.W. (ed.), Israel, W. (ed.): Three hundred years of gravitation*, 676-684 and Preprint - Crnkovic, C. (86,rec.Dec.) 13 p $\bullet$ A.~Ashtekar, L.~Bombelli and O.~Reula,
PRINT-90-0318 (SYRACUSE). 
\bibitem{Thesis-Compere-Seraj-Hajian} 
For reviews e.g. see: 
G.~Compere,
\href{http://arxiv.org/abs/0708.3153}{[arXiv:0708.3153 [hep-th]]} $\bullet$
K.~Hajian,
\href{http://arxiv.org/abs/1508.03494}{[arXiv:1508.03494 [gr-qc]]} $\bullet$
A.~Seraj,
\href{http://arxiv.org/abs/1603.02442}{[arXiv:1603.02442 [hep-th]]}.


\bibitem{Lee-Wald} 
J.~Lee and R.~M.~Wald,
J.\ Math.\ Phys.\  {\bf 31}, 725 (1990) $\bullet$
V.~Iyer and R.~M.~Wald,
Phys.\ Rev.\ D {\bf 50}, 846 (1994)
\href{http://arxiv.org/abs/gr-qc/9403028}{[gr-qc/9403028]}.

\bibitem{Barnich-Brandt} 
G.~Barnich and F.~Brandt,
Nucl.\ Phys.\ B {\bf 633}, 3 (2002)
\href{http://arxiv.org/abs/hep-th/0111246}{[hep-th/0111246]}.


\bibitem{symplectic} 
G.~Comp\`ere, M.~Guica and M.~J.~Rodriguez,
JHEP {\bf 1412}, 012 (2014), \href{http://arxiv.org/abs/1407.7871}{[arXiv:1407.7871 [hep-th]]}.
\bibitem{symplectic-AdS3} 
G.~Comp\`ere, P.~J.~Mao, A.~Seraj and M.~M.~Sheikh-Jabbari,
JHEP {\bf 1601}, 080 (2016)
\href{http://arxiv.org/abs/1511.06079}{[arXiv:1511.06079 [hep-th]]}.

\bibitem{Hajian:2015xlp} 
K.~Hajian and M.~M.~Sheikh-Jabbari,
Phys.\ Rev.\ D {\bf 93}, no. 4, 044074 (2016)
\href{http://arxiv.org/abs/1512.05584}{[arXiv:1512.05584 [hep-th]]}.



\bibitem{HSY-to-come}
 K. Hajian, M. M. Sheikh-Jabbari, and H. Yavartanoo,
Extremal black hole fluff (to be published).

\bibitem{footnote 2}
This is in contrast with the 3D cases where all real values of $J_0$ lead to smooth geometries \cite{AGSY, SY}.

\bibitem{SY} 
M.~M.~Sheikh-Jabbari and H.~Yavartanoo,
Eur.\ Phys.\ J.\ C {\bf 76}, no. 9, 493 (2016), 
\href{http://arxiv.org/abs/1603.05272}{[arXiv:1603.05272 [hep-th]]}.

\bibitem{footnote 3}
Technically, they fall into hyperbolic Virasoro coadjoint orbits \cite{Vir-Orbits, Balog, SY}.


\bibitem{Vir-Orbits}
A.~A.~Kirillov, Funct. Anal. Appl. 15 (2) (1981) 135; Springer Lecture Notes in
Mathematics, vol. 970 (1982) 101 $\bullet$
G.~Segal,
Commun.\ Math.\ Phys.\  {\bf 80}, 301 (1981) $\bullet$
E.~Witten,
Commun.\ Math.\ Phys.\  {\bf 114}, 1 (1988).


\bibitem{Balog} 
J.~Balog, L.~Feher and L.~Palla,
Int.\ J.\ Mod.\ Phys.\ A {\bf 13}, 315 (1998)
\href{http://arxiv.org/abs/hep-th/9703045}{[hep-th/9703045]}.


\bibitem{Maldacena-Maoz} 
J.~M.~Maldacena and L.~Maoz,
JHEP {\bf 0212}, 055 (2002)
\href{http://arxiv.org/abs/hep-th/0012025}{[hep-th/0012025]}.


\bibitem{Carlip-Hardy-Ramanujan} 
S.~Carlip,
Class.\ Quant.\ Grav.\  {\bf 15}, 3609 (1998)
\href{http://arxiv.org/abs/hep-th/9806026}{[hep-th/9806026]}.

\bibitem{BMS-Glenn} 
G.~Barnich and C.~Troessaert,JHEP {\bf 1005}, 062 (2010)
\href{http://arxiv.org/abs/1001.1541}{[arXiv:1001.1541 [hep-th]]};
JHEP {\bf 1112}, 105 (2011)
\href{http://arxiv.org/abs/1106.0213}{[arXiv:1106.0213 [hep-th]]};  
JHEP {\bf 1603}, 167 (2016)
\href{http://arxiv.org/abs/1601.04090}{[arXiv:1601.04090 [gr-qc]]}. 


\bibitem{Compere-Long}
G.~Comp\`ere and J.~Long,
JHEP {\bf 1607}, 137 (2016)
\href{http://arxiv.org/abs/1601.04958}{[arXiv:1601.04958 [hep-th]]}.


\bibitem{DGGP} 
L.~Donnay, G.~Giribet, H.~A.~Gonzalez and M.~Pino,
Phys.\ Rev.\ Lett.\  {\bf 116}, no. 9, 091101 (2016)
\href{http://arxiv.org/abs/1511.08687}{[arXiv:1511.08687 [hep-th]]}; 
JHEP {\bf 1609}, 100 (2016)
\href{http://arxiv.org/abs/1607.05703}{[arXiv:1607.05703 [hep-th]]}.



\bibitem{log-Kerr/CFT} 
A.~Pathak, A.~P.~Porfyriadis, A.~Strominger, O.~Varela,
JHEP {\bf 1704}, 090 (2017)
\href{http://arxiv.org/abs/1612.04833}{[arXiv:1612.04833 [hep-th]]}.


\bibitem{log-sen} 
A.~Sen,
Gen.\ Rel.\ Grav.\  {\bf 44}, 1947 (2012)
\href{http://arxiv.org/abs/1109.3706}{[arXiv:1109.3706 [hep-th]]}.

\bibitem{Hajian:2016kxx} 
 K.~Hajian,
 Gen.\ Rel.\ Grav.\  {\bf 48}, no. 8, 114 (2016)
\href{http://arxiv.org/abs/1602.05575}{ [arXiv:1602.05575 [gr-qc]]}.

\bibitem{Ghodrati:2016vvf}
  M.~Ghodrati, K.~Hajian and M.~R.~Setare,
  Eur.\ Phys.\ J.\ C {\bf 76} (2016) no.12,  701,
 \href{http://arxiv.org/abs/1606.04353}{[arXiv:1606.04353 [hep-th]]}.

\end{thebibliography}
\end{document}